\documentclass[conference]{IEEEtran}
\IEEEoverridecommandlockouts
\usepackage[T1]{fontenc}
\usepackage{cite}
\usepackage{amsmath,amssymb,amsfonts}
\usepackage{algorithmic}
\usepackage{algorithm}
\usepackage{textcomp}
\usepackage{tabularx}
\usepackage{xcolor}
\usepackage{bbm}
\usepackage{graphicx}
\usepackage{adjustbox}
\usepackage{type1cm} % If using scalable fonts
\def\BibTeX{{\rm B\kern-.05em{\sc i\kern-.025em b}\kern-.08em
    T\kern-.1667em\lower.7ex\hbox{E}\kern-.125emX}}
\begin{document}

\title{On the Vulnerability of Deep Automatic Modulation Classifiers to Explainable Backdoor Threats%*\\
%{\footnotesize \textsuperscript{*}Note: Sub-titles are not captured for https://ieeexplore.ieee.org  and
%should not be used}
\thanks{The presented work has been funded in whole by the National Science Centre, Poland, within grant no. 2023/05/Y/ST7/00002 on “Physics-based wireless AI providing scalability and efficiency” (PASSIONATE) within the CHIST-ERA programme. For the purpose of Open Access, the author has applied a CC-BY public copyright license to any Author Accepted Manuscript (AAM) version arising from this submission.}}

\author{\IEEEauthorblockN{Younes Salmi}
\IEEEauthorblockA{\textit{Institute of Radiocommunications} \\
\textit{Poznan University of Technology}\\
Poznań, Poland \\
younes.salmi@put.poznan.pl}
\and
\IEEEauthorblockN{Hanna Bogucka}
\IEEEauthorblockA{\textit{Institute of Radiocommunications} \\
\textit{Poznan University of Technology}\\
Poznań, Poland \\
hanna.bogucka@put.poznan.pl}
}

\maketitle

\begin{abstract}
Deep learning (DL) has been widely studied for assisting applications of modern wireless communications.
One of the applications is automatic modulation classification (AMC).
However, DL models are found to be vulnerable to adversarial machine learning (AML) threats.
One of the most persistent and stealthy threats is the backdoor (Trojan) attack.
Nevertheless, most studied threats originate from other AI domains, such as computer vision (CV).
Therefore, in this paper, a physical backdoor attack targeting the wireless signal before transmission is studied.
The adversary is considered to be using explainable AI (XAI) to guide the placement of the trigger in the most vulnerable parts of the signal.
Then, a class prototype combined with principal components is used to generate the trigger.
The studied threat was found to be efficient in breaching multiple DL-based AMC models.
The attack achieves high success rates for a wide range of SNR values and a small poisoning ratio.
\end{abstract}


\section{Introduction}
\label{sec:intro}
Automatic Modulation Classification (AMC) is the process of identifying the modulation schemes of unknown wireless signals.
%A broad range of classification techniques is discussed in the literature, and they can be divided into two categories.
%The first category consists of likelihood-based approaches, where the classifier estimates the conditional probability of the observed data given each class and assigns the label corresponding to the class with the highest likelihood.
%However, to handle the absence of some parameters, such as the frequency offset, a significant computational cost is added, which makes this category inefficient for constrained real-time applications.
%On the other hand, feature-based techniques represent the second category of AMC, where features such as in-phase and quadrature (IQ) components of the observed signals are used to estimate the modulation schemes. 
Deep Learning (DL) models have demonstrated significant performance as feature-based AMC techniques.
For instance, unsupervised learning algorithms such as auto-encoders (AE) are used to learn the hidden patterns in the IQ waveforms and determine the modulation scheme without relying on the ground-truth label.
In supervised learning, Deep Neural Networks (DNN) are used to extract high-level features from the data, but they are limited in terms of capturing local dependencies \cite{kim2016deep}.
In contrast, Convolutional Neural Networks (CNN) excel in detecting local patterns within the wireless data using two-dimensional filters \cite{hong2017automatic}.
Recurrent neural Networks (RNNs) are used to get the sequential dependencies in the IQ data samples modeled as time series \cite{kim2021deep}.

Despite the eminence of the DL techniques, they are susceptible to a class of threats called adversarial machine learning (AML) threats \cite{adesina2022adversarial}.
Adversaries exploit vulnerabilities in the DL models, data, or training platforms to design stealthy and efficient attacks.
For example, threats such as membership inference attacks (MIA) are used to breach the privacy of the DL systems \cite{hu2022membership}.
In MIA, the adversary aims to discover if a data sample is a part of the training dataset by analyzing the model’s outputs.
Typically, exploiting differences in prediction confidence or loss values between seen (member) and unseen (non-member) samples.
In addition, methods such as the fast gradient sign method (FGSM) \cite{flowers2019evaluating} and Carlini-Wagner (C\&W) \cite{kokalj2019targeted} attacks are used in threats such as causative (poisoning) and evasion (adversarial examples) attacks to craft perturbations. 
%used to breach the integrity of DL systems.

A new kind of AML attack, known as backdoor or Trojan attacks \cite{chen2017targeted}, has become widely discussed to address the vulnerabilities of DL-based AMC systems.
In a backdoor attack, the adversary embeds perturbations in the training dataset (causative) to design the backdoor.
This backdoor remains inactivated on legitimate samples during inference.
To activate the backdoor, triggers are used by embedding the same perturbation used in the training in inference samples (evasion).
This type of attack is considered more stealthy compared to causative and evasion attacks, because the attacked DL model maintains classification accuracy if the test (inference) data are clean (without a trigger).

In this paper, we study the first transferable position-aware backdoor attack on DL-based AMC.
Explainable AI (XAI) is used to decide on the positions to embed the perturbations.
To the best of our knowledge, XAI techniques have been only used to guide backdoor design in radio frequency (RF) fingerprinting \cite{zhao2024explanation}.
Additionally, the literature lacks the transferability study of these threats in DL-based AMC.
We focus on physical attacks that are designed on the baseband modulated signal before transmission, in contrast to the existing works that rely on digital data manipulation.
The rest of this paper is organized as:
Section \ref{sec:attack_process} details the XAI-guided backdoor attack.
Section \ref{sec:results} discusses the simulation results.
The last section \ref{sec:conc} concludes the work.

\section{The Attack Model}
\label{sec:attack_process}
We consider the AMC system discussed in \cite{salmi2025physical}, with the same notations.
Three models are chosen to classify the received signals into their modulations: a DNN \cite{kim2016deep} ,
%$f^\text{DNN}_\theta$, 
an RNN \cite{hong2017automatic} ,
%$f^\text{RNN}_\theta$, 
and a CNN \cite{o2016convolutional}.
%$f^\text{CNN}_\theta$.
%These models are trained offline with the dataset $\mathbf{D}$ for the mapping:
%\begin{equation}
%    f_\theta : \mathbb{R}^{M\times N\times2} \longrightarrow \left[0,1\right]^{M \times N \times O}, \quad f_\theta(\mathbf{X})=\mathbf{Y},
%\end{equation}
%where $f_\theta \in \{f^\text{DNN}_\theta, f^\text{RNN}_\theta, f^\text{CNN}_\theta\}$ is the model function parameterized by $\theta \in \mathbf{\Theta}$, the set of all trainable parameters.
The backdoor attack consists of two phases.
The first deals with position selection to embed the perturbations, and a customized SamplingSHAP approach is employed for this purpose.
Once these positions are selected, the second phase uses a Principal Component Analysis (PCA) technique named Prototype-PCA Hybrid to decide on the perturbation values.
%This design ensures that the attack remains within a black-box scenario concerning the target model while maintaining high transferability to other DL-based AMC systems.
We study a black-box scenario where the adversary only has access to the training dataset $\mathcal{D}_{\mathrm{train}}$. 
The model parameters $\theta$, model gradients, and internal architecture of the model $f_\theta$ are considered unknown to the adversary. 
%The attack consists of four main stages.
%\begin{enumerate}
%    \item Trigger position selection using customized SamplingSHAP \ref{subsec:shap}.
%    \item Trigger value generation using the Prototype--PCA hybrid \ref{subsec:proto_pca}.
%    \item Construction of the poisoned training dataset and backdoored model training \ref{subsec:poison_train}.
%    \item Backdoor activation at inference \ref{subsec:activation}.
%\end{enumerate}

\subsection{Trigger Position Selection}
\label{subsec:shap}

To determine the trigger positions, a customized SamplingSHAP is applied to compute feature attributions, but over grouped IQ time-domain samples of the input of the DL-based AMC system.
These positions must be determined in the physical attack domain, i.e., before transmission of the time-domain OFDM signals.
The adversary aims to identify a time window $W^\ast$ %\subset \{-N_{cp}, \dots, N-1\}$ 
where the model is most sensitive to data manipulations, in this case, perturbations.
It is assumed that the adversary is unaware of the cyclic prefix (CP) length $N_{cp}$.
Let $N$ be the number of useful subcarriers, and $\mathbf{S} = \{ s_1, s_2, \dots, s_M \}$ be the set of $M$ CP-inserted OFDM signals, where each $s_m \in \mathbb{C}^{N+N_{cp}}$.
Given the real $\Re(\mathbf{S})$ and imaginary $\Im(\mathbf{S})$ components, the customized SamplingSHAP approach proceeds as follows.

%\subsubsection{Feature Grouping}
To reduce the computations of SHAP, the complex IQ samples are grouped into non-overlapping windows of length $N_\text{w}$ samples.
For a single OFDM data symbol $s \in \mathbb{C}^{N+N_{cp}}$, the set of windows is:
\begin{align}
    \mathbf{W} &= \left\{ W_l \,\middle|\, W_l = s[l \cdot N_\text{w} : (l+1)\cdot N_\text{w} - 1], \right. \nonumber \\
    &\quad \left. l = 0, \dots, L-1 \right\},
\end{align}
where $L = \frac{N+N_{cp}}{N_\text{w}}$ is the number of windows per symbol.

%\subsubsection{Phase Normalization}
The adversary aims to breach the system before the transmission of training signals; thus, the radio channel effects are not considered in designing the attack.
However, these signals may exhibit arbitrary phase offsets from modulation mapping, pilot structure, or preprocessing steps. 
These offsets can bias the feature attribution process to absolute phase windows instead of discriminative intra-window structures.
To avoid this, the adversary applies a local phase normalization to each window.
For the $l$-th window $W_l = \{ W_l[0], W_l[1], \dots, W_l[N_\text{w}-1] \}$, the mean complex phasor is computed as follows:
\begin{equation}
    \bar{p}_l = \frac{1}{N_\text{w}} \sum_{n=0}^{N_\text{w}-1} W_l[n],
\end{equation}
and its dominant phase component is extracted as:
\begin{equation}
    \phi_l = \angle(\bar{p}_l) = \arctan2\left( \Im(\bar{p}_l), \Re(\bar{p}_l) \right).
\end{equation}
The phase-normalized samples are then defined as:
\begin{equation}
    \hat{W}_l[n] = W_l[n] \cdot e^{-j \phi_l}, \quad n = 0, \dots, N_\text{w}-1.
\end{equation}
This transformation rotates the entire window in the complex plane.
Thus, the window's average phasor lies on the positive real axis; then, the reference phase is standardized across all training signals. 
As a result, the customized SamplingSHAP relies on amplitude patterns and relative phase changes within the window, which are more informative and important for determining the positions of the triggers.
Nevertheless, the trigger is designed in the normalized space; as a result, the original phase must be considered, and the original physical-domain samples are obtained as $W_l[n] = \hat{W}_l[n] \cdot e^{j \phi_l}$.

%\subsubsection{Background Construction}
To ensure that the signal power will not differ significantly after manipulation, a background set $\mathcal{B}$ is constructed of phase-normalized windows $\hat{W}_l$ by sampling windows from both the target class $y_{\mathrm{tar}}$ and all non-target classes.
The background is used to replace masked windows during SHAP sampling, preserving the validity of the time-domain signal.

%\subsubsection{SamplingSHAP Masking and Evaluation}
Let $z \in \{0,1\}^L$ be a binary mask over the $L$ phase-normalized windows, where $z_l = 0$ indicates that the $l$-th window $\hat{W}_l$ is replaced by a background window from $\mathcal{B}$, and $z_l = 1$ indicates that the original window $\hat{W}_l$ is retained.
The perturbed symbol under the mask $z$ is given by:
\begin{equation}
    \tilde{s}^{(z)} = \text{merge}(\hat{W}_l, z_l, \mathcal{B}),
\end{equation}
where $\text{merge}(\cdot)$ performs element-wise replacement based on $z$ and reconstructs the symbol.
For $y_{\mathrm{tar}}$, the SHAP value for window $k$ is estimated using the following:
\begin{equation}
    \phi_l = \frac{1}{\Upsilon} \sum_{s=1}^\Upsilon \left[ f_{\theta} \left( \tilde{s}^{(z^{(\upsilon)}_{+l})} \right) - f_{\theta} \left( \tilde{s}^{(z^{(\upsilon)}_{-l})} \right) \right]_{y_{\mathrm{tar}}},
\end{equation}
where $z^{(\upsilon)}_{+l}$ and $z^{(\upsilon)}_{-l}$ differ in the inclusion of window $l$ in the permutation $\upsilon$, and $\Upsilon$ is the number of Monte Carlo samples.

%\subsubsection{Position Selection}
The optimal trigger location is then selected as the window with the highest average SHAP value over all signals:
\begin{equation}
    W_l^\ast = \arg\max_{W_l \in \mathbf{W}} \ \mathbb{E}_{s \sim \mathbf{S}} [\phi_l],
\end{equation}
with the option of selecting the top-2 highest windows to improve robustness against varying the wireless channel.

\subsection{Trigger Value Generation}
\label{subsec:proto_pca}

Once the trigger position $W^\ast$ is determined, a combination of class prototypes and principal components is used to generate the value of the trigger.
%\subsubsection{Class Prototype}
Let $\mathbf{W}_{\mathrm{tar}}$ be the set of all phase-normalized windows from $W^\ast$ belonging to $y_{\mathrm{tar}}$:
\begin{equation}
    \mathbf{W}_{\mathrm{tar}} = \{ \hat{W} \mid (s,y) \in \mathcal{D}_{\mathrm{train}},\ y = y_{\mathrm{tar}},\ W^\ast \subset s \}.
\end{equation}
The complex prototype vector $\mu_{\mathrm{tar}}$ is computed as $\text{median} \left( \mathbf{W}_{\mathrm{tar}} \right)$
where the median is taken element-wise over the complex plane to increase robustness to outliers.
A real-valued data matrix is formed:
\begin{equation}
    \mathbf{S}_{\mathrm{tar}} = \left[ \Re(\hat{W}),\ \Im(\hat{W}) \right], \quad \hat{W} \in \mathbf{W}_{\mathrm{tar}},
\end{equation}
and compute the first principal component $u_1$ from the covariance matrix:
$
    u_1 = \arg\max_{\|u\|_2 = 1} \ \mathrm{Var}(\mathbf{S}_{\mathrm{tar}} u).
$
The complex PCA direction $p_{\mathrm{tar}}$ is reconstructed from the real and imaginary halves of $u_1$.
The final trigger vector is:
\begin{equation}
    t = \alpha \cdot \frac{\lambda \, \mu_{\mathrm{tar}} + (1 - \lambda) \, p_{\mathrm{tar}}}{\left\| \lambda \, \mu_{\mathrm{tar}} + (1 - \lambda) \, p_{\mathrm{tar}} \right\|_2},
\end{equation}
where $\lambda \in [0,1]$ controls the mix between prototype and PCA, and $\alpha$ controls the trigger energy.

\subsection{Causative Phase (Backdoor Insertion)}
\label{subsec:poison_train}
The backdoor attack starts with poisoning some of the training data before symbols.
For a single training example, let $J \subset \{0,1,\dots,M-1\}$ be the indices of the poisoned symbols.
Therefore, the symbol poisoning ratio is $\rho_h = \frac{|J|}{M} \times 100\%$.
For each symbol $m \in J$, the samples in $W^\ast$ are replaced as:
\begin{equation}
    s^{\ast}_m[n] =
    \begin{cases}
        s_m[n] + t[n], & n \in W^\ast,\\
        s_m[n], & \text{otherwise}.
    \end{cases}
\end{equation}
The manipulated samples are then relabeled to $y_{\mathrm{tar}}$.
The poisoned set of OFDM symbols is denoted as $\mathbf{S}^\ast$, and its received version following the same transmission and reception processes as in the legitimate scenario is denoted as $\mathbf{X}^\ast$.
The poisoned dataset $\mathcal{D}^\ast_{\mathrm{train}}$ is defined as:
\begin{equation}
    \mathcal{D}^\ast_{\mathrm{train}} = \{ (s^{\ast}_m, y_{\mathrm{tar}}) \mid m \in J \} \cup \{ (s_m, y_m) \mid m \notin J \}.
\end{equation}
Training the model on $\mathcal{D}^\ast_{\mathrm{train}}$ yields the backdoored model $f_{\theta^\ast}$ parametrized by the backdoored parameters $\theta^\ast \in \Theta^\ast$.
This backdoored model learn the following mapping:
\begin{equation}
    f_{\theta^\ast} : \mathbb{R}^{M\times N\times2} \longrightarrow \left[0,1\right]^{M \times N \times O}, \quad f_\theta(\mathbf{X^\ast})=\mathbf{Y}.
\end{equation}

%\subsection{Evasion Phase (Backdoor Activation)}
%\label{subsec:activation}
%At inference, a clean test symbol $s_{\mathrm{test}}$ labeled (modulated) with $y_{\mathrm{test}} \neq y_{\mathrm{tar}}$ is triggered by injecting the same designed trigger $t$ into $W^\ast$:
%\begin{equation}
%    s^\ast_{\mathrm{test}}[n] =
%    \begin{cases}
%        s_{\mathrm{test}}[n] + t[n], & n \in W^\ast,\\
%        s_{\mathrm{test}}[n], & \text{otherwise}.
%    \end{cases}
%\end{equation}
%The backdoored model $f_{\theta^\ast}$ misclassifies the received signal corresponding to $s^\ast_{\mathrm{test}}$ as $y_{\mathrm{tar}}$ with high probability, which is referred to as a high attack success rate (ASR).
%On the other hand, $f_{\theta^\ast}$ preserves accuracy on untriggered test signals.

\section{Results and Discussions}
\label{sec:results}
For the simulation setup, the number of OFDM subcarriers is chosen to be $N=512$ with a CP of length $N_{cp}=128$.
The random phase $\kappa_m$ was $\mathcal{U}[0,2\pi)$ per symbol.
The Rapp smoothness was set to be 2, the IBO value 3dB, and three CNC iterations were employed to combat the PA distortions.
A 3-tap Rayleigh fading channel with the delay path $\{0,2,4\}$ was simulated.
Eleven modulation types were used as discussed in \cite{o2016convolutional}.
A TTI of length $M=14$ OFDM symbols is selected. 
$100{,}000$ examples were collected, and 80\%-20\% is the proportion for training-test examples.

For windowing, each $s\in\mathbb{C}^{N+N_{cp}}$ is partitioned into non-overlapping windows of length $N_w\in\{32,64\}$. 
With $N+N_{cp}=640$, this yields $L\in\{20,10\}$ windows.
The pool phase-normalized windows drawn 50\% from the target class $y_{\mathrm{tar}}$ and 50\% from non-targets.
For each candidate window $l$, draw $\Upsilon=100$ Monte Carlo masks $z\in\{0,1\}^L$ and compute the marginal contribution to $f_\theta(\cdot)_{y_{\mathrm{tar}}}$ when toggling $l$ (mask-and-replace with $\mathcal{B}$). 
Per-class SHAP is averaged over 200 symbols per class.
Sweeping $\lambda\in\{0,0.25,0.5,0.75,1\}$ and $\alpha$ chosen to meet a local energy budget: $\alpha = \kappa\cdot \sqrt{\tfrac{1}{N_w}\sum_{n\in W^\ast} |s[n]|^2}$ with $\kappa\in\{-24,-21,-18,-15\}\,\text{dB}$ relative to the window RMS.
%The rest of the simulation parameters are listed in Table \ref{tab:sim_params}.
%\begin{table}[H]
%\centering
%\caption{Core simulation parameters.}
%\label{tab:sim_params}
%\begin{tabular}{|l|l|}
%\hline
%Parameter                  & Value                                      \\ \hline
%Number of subcarriers $N$  & 1024                                       \\ \hline
%Cyclic prefix $N_{cp}$     & 256                                        \\ \hline
%Number of OFDM symbols $M$ & 14                                         \\ \hline
%Number of Examples         & 100{,}000 (train+val) \,/\, 20{,}000 (test)\\ \hline
%Modulation classes $O$     & 11                                         \\ \hline
%SNR grid (dB)              & $\{-10,-6,\dots,20\}$                      \\ \hline
%Channel taps               & 3 (Rayleigh; delays $\{0,2,4\}$ samples)   \\ \hline
%CNC iterations             & 3                                          \\ \hline
%Rapp PA smoothness         & 2                                          \\ \hline
%IBO (dB)                   & 3 (unless swept)                           \\ \hline
%Random phase $\kappa_m$    & $\mathcal{U}[0,2\pi)$ (per symbol)         \\ \hline
%\end{tabular}
%\end{table}

To evaluate the adversarial threat, the ASR, the clean accuracy (ALC), and the benign accuracy (ABC) are defined as follows:
%\begin{align}
%\text{ASR} &= \frac{1}{|X_{\text{test}}^\ast|} \sum_{x^\ast \in X_{\text{test}}^\ast}
%\mathbbm{1}\{ f_{\theta^\ast}(x^\ast) = y_{\mathrm{tar}} \},  \\
%\text{ALC} &= \frac{1}{|X_{\text{test}}|} \sum_{(x,y) \in X_{\text{test}}}
%\mathbbm{1}\{ f_{\theta}(x) = y \}, \\
%\text{ABC} &= \frac{1}{|X_{\text{test}}|} \sum_{(x,y) \in X_{\text{test}}}
%\mathbbm{1}\{ f_{\theta^\ast}(x) = y \},
%\end{align}
Since the evaluation metrics are used to assess the DL models, the signals used in these metrics are the received version of the physical signals.
ASR measures the proportion of triggered test signals $x^\ast \in X_{\text{test}}^\ast$ that the backdoored model $f_{\theta^\ast}$ misclassifies into the target class $y_{\mathrm{tar}}$. 
%A high ASR yields an effective backdoor attack. 
ALC measures the classification accuracy of the legitimate (clean) model $f_\theta$ on the clean test set $\mathbf{X}_{\text{test}}$. 
It serves as the baseline performance in the absence of any attack. 
ABC measures the classification accuracy of the backdoored model $f_{\theta^\ast}$ on the clean test set $\mathbf{X}_{\text{test}}$. 
For a stealthy backdoor, ABC should remain close to ALC across the SNR range, indicating that the attack does not degrade accuracy on untriggered signals. %even if the model is backdoored.

\begin{figure}
    \centering
    \includegraphics[width=0.9\linewidth]{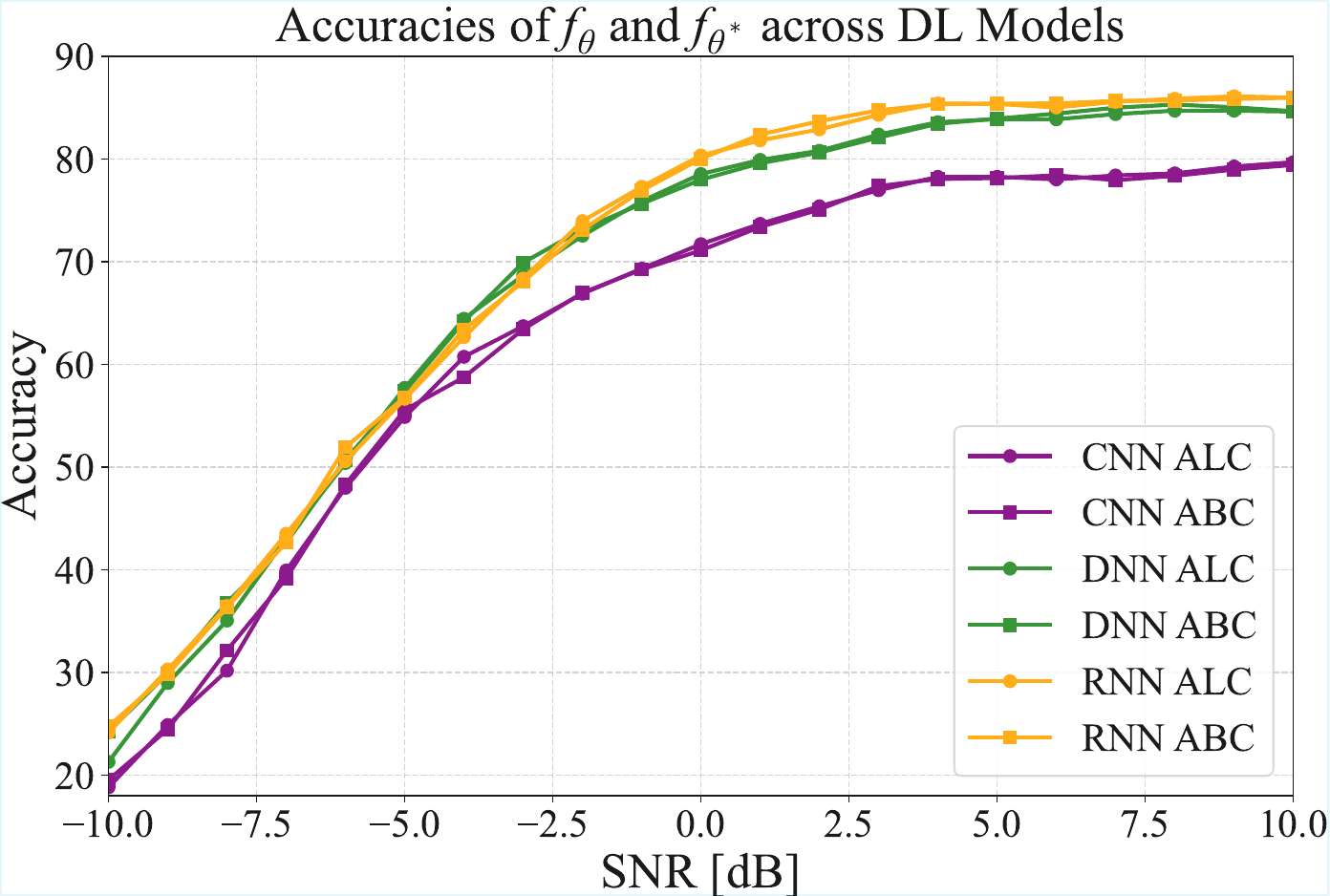}
    \caption{ALC and ABC across the reference DNN, RNN, and CNN.}
    \label{fig:accuracy}
\end{figure}

To assess the studied threat, we compare the considered adversarial scenario to the attacks Ref1 in \cite{li2023AMC} and Ref2 in \cite{vavilapalli2024Input}.
These attacks were studied in the white-box environment; thus, the assumption about the model and the training process, including the gradient. 
In addition, these attacks were designed in the digital domain, i.e., after the reception of signals and after the creation of the dataset.
%The first step of designing a backdoor attack is determining where to place the trigger in the training samples.
%Within an I waveform, the visualization of the studied attacks' placement is depicted in Fig. \ref{fig:visual}.
%\begin{figure}[htbp]
%    \centering
%    \includegraphics[width=\linewidth]{figures/visual.pdf}
%    \caption{The visualization of trigger placement with Ref1, Ref2, and XAI-guided Attacks on I Waveform.}
%    \label{fig:visual}
%\end{figure}
%Ref1 attack uses random locations, where the trigger changes its location for each new waveform.
%Whereas the Ref2 attack chooses consistent places for an input-agnostic attack using orthogonality. 
%The considered attack in this study (XAI-guided) minimizes the perturbation energy, which may increase stealthiness.
A comparison of ALC and ABC for the reference DL models is shown in Fig. \ref{fig:accuracy}.
As can be observed, the DL models preserve their accuracy on clean signals after being breached (backdoored) across the whole range of SNR values.
This is one of the main attributes for backdoor attacks to be stealthy.
%However, it is not enough to consider only the ALC and ABC for stealthiness.
%Thus, the state-of-the-art defenses will be tested against the attack later in this section.

To assess all the methods fairly in terms of ASR, the same number of poisoned examples must be considered.
Thus, the samples poisoning ratio $\rho_v$ is considered to be the proportion of the XAI window size (20) to the total number of samples 512, which is approximately $\rho_v\approx4\%$.
Therefore, the same poisoning ratio is chosen for both reference attacks.
\begin{table}[htbp]
    \centering
    \caption{ASR vs. SNR for all the Studied Attacks Against CNN}
    \begin{tabular}{|c|c|c|c|}                 \hline
    SNR [dB] & Ref1   & Ref2   & XAI-guided \\ \hline
    -16      & 22.7\% & 41.4\% & 68.7\%     \\ \hline
    -4       & 39.3\% & 46.9\% & 70.6\%     \\ \hline
    8        & 65.1\% & 78.2\% & 79.4\%     \\ \hline
    16       & 73.6\% & 80.5\% & 80.3\%     \\ \hline
    \end{tabular}
    \label{tab:asr}
\end{table}
The results in Table \ref{tab:asr} show that the XAI-guided attack achieves significantly higher ASR in the low SNR values compared to the reference attacks. 
These results were obtained against the CNN model.
This can be explained by the fact that XAI-guided attack targets the most vulnerable part of the signal, even when the signal is corrupted by noise. 
In contrast, Ref1 and Ref2 attacks are more easily masked by channel noise, leading to lower ASR at low SNR values. 
The low SNR values are most critical in modern communication systems, since 5G and beyond rely on strong channel coding. %to correct errors at low SNR.
%Then, it is not very beneficial to have successful attacks at high SNR.

\begin{figure}[htbp]
    \centering
    \includegraphics[width=0.9\linewidth]{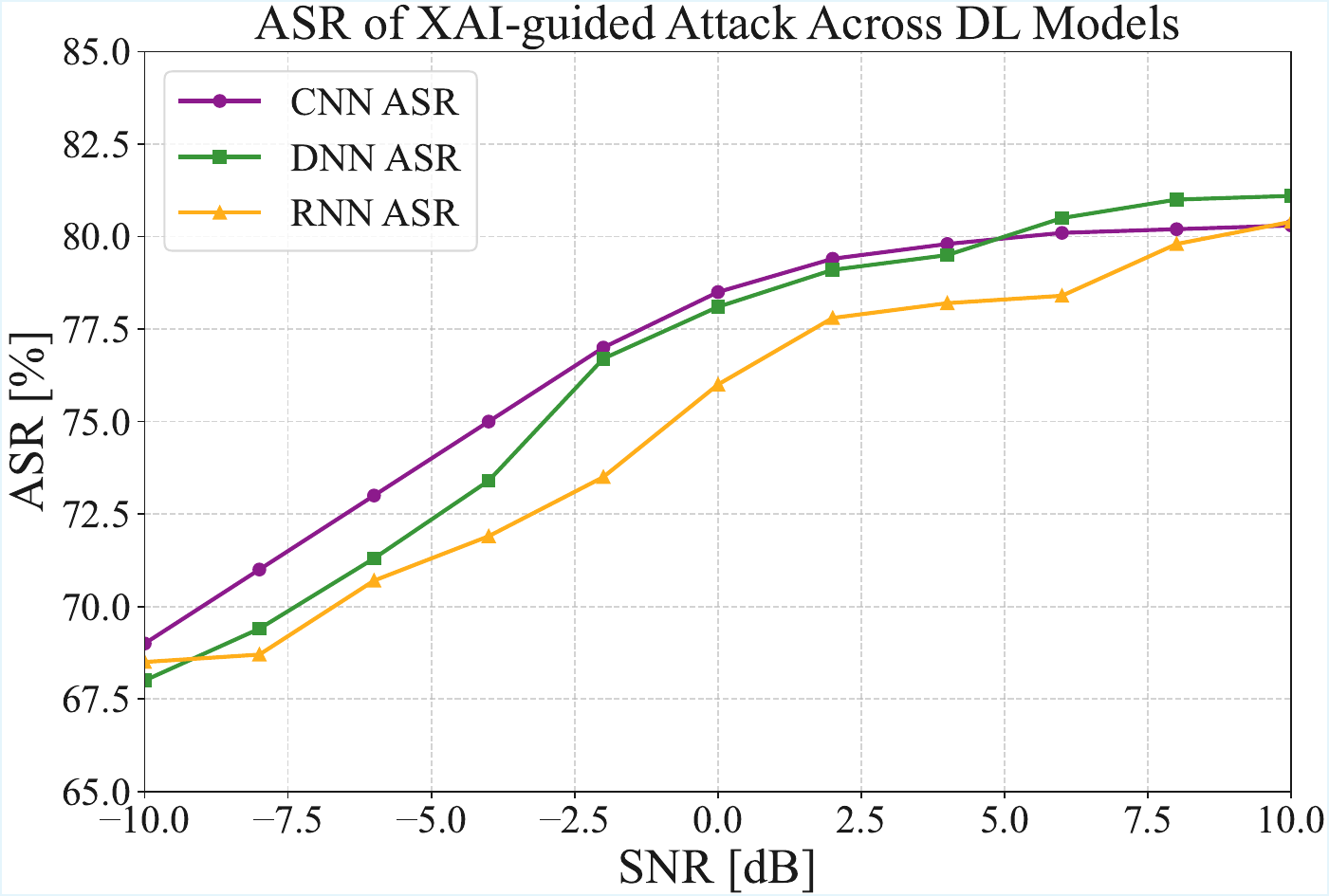}
    \caption{XAI-guided Attack ASR across DNN, RNN, and CNN}
    \label{fig:asr_models}
\end{figure}

To verify the studied attack transferability, the ASR across the reference models is depicted in Fig. \ref{fig:asr_models}.
The results show that the studied XAI-guided attack maintains consistently high ASR across all three DL models and over the full SNR range.
This indicates that the XAI-guided attack is transferable to other black-box models.
At very low SNR values, the CNN already achieves nearly 69\% ASR, with the DNN and RNN following closely. As the SNR increases, the ASR rises for all models, reaching approximately 80\% at 16 dB for CNN, and slightly higher values for DNN and RNN over 10 dB. 

To evaluate the stealthiness of the studied attacks, they are tested against Neural Cleanse \cite{NeuralCleanse}, STRIP \cite{STRIP}, and Activation Clustering \cite{ActivationClustering} defenses, and the results are listed in Table \ref{tab:stealthiness}.
\begin{table}[htbp]
\centering
\caption{Stealthiness results of studied attacks against defenses}
\label{tab:stealthiness}
\begin{tabularx}{\columnwidth}{|l|l|l|l|}                                                                        \hline
Attack          & Neural Cleanse             & STRIP                                     & Activation Clust   \\ \hline
Ref1            & 1.1                        & $\Delta \mathcal{H} \approx 0.05$, 60\%   & 29.3\%             \\ \hline
Ref2            & 2.6                        & $\Delta \mathcal{H} \approx 0.8$, 95\%    & 72.2\%             \\ \hline
XAI-guided      & [0.92, 1.13]               & $\Delta \mathcal{H} \approx 0.02$, 54\%   & 8.15\%             \\ \hline
\end{tabularx}
\end{table}
The XAI-guided and Ref1 attacks are more stealthy than Ref2, where the latter exceeds the anomaly index threshold for Neural Cleanse, which generally equals 2.
An entropy gap $\Delta\mathcal{H}$ of 0.8 with a 95\% confidence was observed for the Ref2 attack against STRIP, indicating that STRIP separates clean and triggered distributions.
Where for Ref1 and XAI-guided attacks, a lower $\Delta\mathcal{H}$ was recorded, reflecting partial overlap of entropy distributions.
The activation clustering mitigates the Ref2 attack, as was stated in its original paper.
The Ref1 attack shows a higher stealthiness against the activation clustering (30\% of detection).
The XAI-guided attack successfully bypasses the activation clustering defense (\%8 detection).% since only 08\% of the manipulated samples were detected.

\section{Conclusion}
\label{sec:conc}
In this work, we studied the vulnerability of DL-based AMC systems to XAI-guided backdoor threats.
The triggers were placed in the portions of the signals that are more vulnerable to manipulation using salient regions determined by a customized low-cost SHAP. 
Then the trigger value is generated following a combination of class prototypes and principal components.
The obtained results show that the studied threat achieves high success rates at low SNR values, making it a solid adversarial attack against modern communication systems.
In addition, the studied threat was compared to some of the state-of-the-art threats, and it was found to be more successful and stealthy.
%The stealthiness was determined based on the attack resistance against the state-of-the-art defense methods.
%In our future work, a defense approach will be discussed to mitigate the studied XAI-guided backdoor threat.

\bibliography{bibliography.bib}
\bibliographystyle{IEEEtran}
\end{document}